# Constant Time Quantum search Algorithm Over A Datasets: An Experimental Study Using IBM Q Experience

Kunal Das[1*], Arindam Sadhu[1#]

[1]Narula Institute of Technology, 81, Nilganj Road,Agrapara,kol-700109. India

[*]kunal.das@nit.ac.in, [#]arindam.hit11@gmail.com

**Abstract-** In this work, a constant time Quantum searching algorithm over a datasets is proposed and subsequently the algorithm is executed in real chip quantum computer developed by IBM Quantum experience (IBMQ). QISKit, the software platform developed by IBM is used for this algorithm implementation. Quantum interference, Quantum superposition and $\pi$ phase shift of quantum state applied for this constant time search algorithm. The proposed quantum algorithm is executed in QISKit SDK local backend 'local_qasm_simulator', real chip 'ibmq_16_melbourne' and 'ibmqx4' IBMQ. Result also suggest that real chip ibmq_16_melbourne is more quantum error or noise prone than ibmqx4.

## 1. Introduction:

After a decade of exhaustive research on Quantum computing, it becomes a reality and extensive focus is given by researcher and industry. Exponential time problems or super polynomial time problems can be solvable in polynomial time using Quantum computing i.e. all NP problems can be solvable in polynomial time. The reduced polynomial time solvable Quantum algorithm development is an emerging field of research in Quantum Computing (QC). In the year of 2017, IBM Quantum Experience (IBMQ) the QC research initiative first lunch 5 Qubits Quantum computer in cloud service [1-2]. IBM QISKit is a software platform developed by IBM, accelerates the QC research interest. The dream of Richard Feynman and Toffoli is becoming reality[3-4]. The superposition, interference and entanglement are very interesting and spooky phenomenon. In Quantum computer quantum bits or qubits is primary unit and unlike classical bits or cbits. Qubit can be represented as $|0\rangle$, $|1\rangle$ and superposition of $|0\rangle$ and $|1\rangle$. The superposition of $|0\rangle$ and $|1\rangle$ can be represented as $|\psi\rangle = \alpha|0\rangle + \beta|1\rangle$, where $\alpha, \beta$ are two complex numbers such that $|\alpha^2| + |\beta^2| = 1$. where $|\alpha^2|$ and $|\beta^2|$ are the probability to be $|0\rangle$ and $|1\rangle$ respectively. if $|\alpha^2| = 0$ the quantum state $|\psi\rangle = |1\rangle$ and if $|\beta^2| = 0$ the quantum state $|\psi\rangle = |0\rangle$ i.e. the probability of quantum state to become $|1\rangle$ and $|0\rangle$ is 1 respectively and otherwise $|\psi\rangle$ represent the superposition of $|0\rangle$ and $|1\rangle$. The entangled state of n qubits produce equivalent $2^n$ cbits computation. Entanglement in quantum computation empowered the computational aspect i.e. n qubits computer is equivalent to $2^n$ cbit classical computer. In the year of 1996, Lov. K. Grover first proposed a $O(\sqrt{N})$ Quantum search algorithm to search a over unsorted data set by means of amplitude amplification technique [5]. In 2008, Ahmed Younes demonstrated constant time quantum search algorithm over unstructured list to get yes/no answer with certainty [6].

In this research work, we proposed a constant running time quantum search algorithm for unique match over random datasets. The proposed algorithm is executed in IBMQ quantum experience QISKit software platform. The proposed quantum circuit is executed in real chip ibmq_16_melbourne and ibmqx4 quantum computer using cloud service[1-2].

Rest of the paper is organized as follows, the Section 2, demonstrate the basic Quantum gates. In section 3, related work is explored. The proposed algorithm is reported in section 4. The simulation, execution and validation of proposed algorithm is demonstrated in section 5. Finally, the conclusion is drawn in section 6.

## 2. Basic Quantum Gates

*a. CNOT or control NOT:* CNOT is a basic Quantum gate in quantum computing. In the operational principle, the CNOT gate flips the second the quantum bit(termed as target bit), if first quantum bit(known as control bit) is $|1\rangle$, otherwise it will remain same. The operation of a 2-qubit CNOT gate is described as shown at equation 1 [7-9].

$$|00\rangle \rightarrow |00\rangle, \ |01\rangle \rightarrow |01\rangle, \ |10\rangle \rightarrow |01\rangle, \ |11\rangle \rightarrow |10\rangle \qquad (1)$$

The unitary matrix representation of a CNOT gate is given by

$$CNOT = \begin{bmatrix} 1 & 0 & 0 & 0 \\ 0 & 1 & 0 & 0 \\ 0 & 0 & 0 & 1 \\ 0 & 0 & 1 & 0 \end{bmatrix}$$

*b. Pauli's gate:* Pauli's gate or operator is a fundamental gate and it has great importance in quantum computing as a single bit operator. Four types of Pauli's operator are there: identity operator, X operator, Y operator and Z operator. In this section all types are discussed excluding identity operator [7-9]. The computational basis of individual operators is following.

(i) X Gate: This gate is denoted by $\sigma_1$ or $\sigma_x$ or X. Its operational principle is same as NOT gate. It acts as follow.

$$X|0\rangle = |1\rangle \quad and \quad X|1\rangle = |0\rangle$$

Functional block diagram of X gate is expressed in following figure 1.

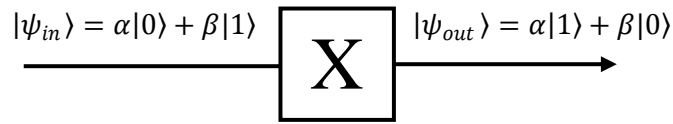

Fig 1. Block diagram of X gate

(ii) Y gate:- The next Pauli's operator is Y gate. The computational basis of Y operator is following. It can be denoted as $\sigma_2$ or $\sigma_y$ or Y.

$$Y|0\rangle = -i|1\rangle \quad and \quad Y|1\rangle = i|0\rangle$$

Functional block diagram of Y gate is expressed in following figure 2.

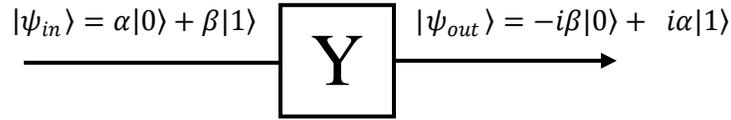

Fig 2. Block diagram of Y gate

(iii) Z gate:- Final Pauli's operator is called Z gate. It can be denoted as $\sigma_3$ or $\sigma_z$ or Z. Some time it is also called phase flip gate as phase difference between input and output is $\pi$ [7-9]. Computational process and functional block diagram are following.

$$Z|0\rangle = |0\rangle \quad and \quad Z|1\rangle = -|1\rangle$$

Functional block diagram of Z gate is expressed in following figure 3.

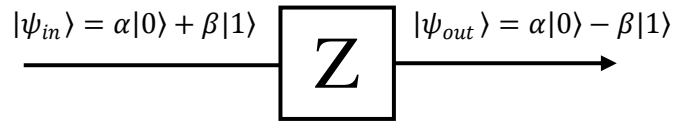

Fig 3. Block diagram of X gate

2×2 matrix form of X, Y and Z operator are mentioned are following.

$$X = \begin{bmatrix} 0 & 1 \\ 1 & 0 \end{bmatrix}, \quad Y = \begin{bmatrix} 0 & -i \\ i & 0 \end{bmatrix}, \quad Z = \begin{bmatrix} 1 & 0 \\ 0 & -1 \end{bmatrix}$$

*c. Phase Shift Gate:* Another very important operator is phase shift gate in quantum computing [7-9]. More over generally shift gate is expressed in following matrix form.

$$P = \begin{bmatrix} 0 & 1 \\ 1 & e^{i\theta} \end{bmatrix}$$

As this gate shift the relative phase of amplitude $\alpha$ and $\beta$, so it is known as phase shift gate. Z gate, S gate and T gate are the example of phase shift gate, where $\theta$ is $\pi$, $\pi/2$, and $\pi/4$ respectively.

*d. Hadamard gate:* Hadamard gate, denoted by H is one of the prime member in quantum computing family.. In computational basis, the hadamard gate has turned to state that had probability $|\alpha^2|$ of a systems finding state $|0\rangle$ into a state that has probability $|\alpha^2|$ of finding the system in state $|+\rangle$ and had probability $|\beta^2|$ of a systems finding state $|0\rangle$ into a state that has probability $|\beta^2|$ of finding the system in state $|-\rangle$. The hadamard gate is represented as [7-9]

$$H = \frac{1}{\sqrt{2}}\begin{bmatrix} 1 & 1 \\ 1 & -1 \end{bmatrix}$$

And  $H|0\rangle = \frac{1}{\sqrt{2}}(|0\rangle + |1\rangle)$     and     $H|1\rangle = \frac{1}{\sqrt{2}}(|0\rangle - |1\rangle)$

## 3. Related works

In 1996, Lov. K. Grover, first introduced a quantum search algorithm and formally known as Grover's search algorithm[5]. In this works, Grover described that in a telephone directory, there are N numbers of telephone numbers and telephone numbers are ordered completely random in order. The searching of one telephone numbers in classical algorithm required $O(N)$ running time. Grover has shown that in Quantum algorithm running time required $O(\sqrt{N})$. In Grover's algorithm, along with quantum interference and superposition of state the amplitude amplification technique is utilized to increase the amplitude of desire quantum state i.e. search key $(x_0)$.

In this algorithm [5, 10-11], the first step of algorithm is started with 'n' numbers of quantum register of 'n' qubits, where 'n' is required to specify $N=2^n$ numbers of search space and all 'n' qubits are initialized to $|0\rangle$. In step 2, The 'n' Parallel Hadamard gate is applied each 'n' qubits. Resultant will caused a quantum interference among 'n' qubits and it is called as Hadamard transformation. The Hadamard transformation is represented by

$$H^{\otimes n}|x\rangle = \frac{1}{\sqrt{2^n}}\sum_{y=0}^{2^n-1}(-1)^{x.y}|y\rangle \qquad (2)$$

where $x.y = \sum_{i=0}^{2^n-1} x_i y_i$ is qubit wise AND of $x_i$ and $y_i$ where $x_i$ and $y_i$ are either 0 or 1. Now example let us assume n=2 $|x\rangle = |11\rangle$ and $H^{\otimes 2}|11\rangle = (H \otimes H)|11\rangle$ It can be simplify to $H^{\otimes 2}|11\rangle = (H|1\rangle)(H|1\rangle)$ further it can be simplify to $H^{\otimes 2}|11\rangle = \frac{1}{\sqrt{2^2}}(|0\rangle - |1\rangle)(|0\rangle - |1\rangle) = \frac{1}{\sqrt{2^2}}(|00\rangle - |01\rangle - |10\rangle + |11\rangle)$ and similarly, $H^{\otimes 2}|00\rangle = \frac{1}{\sqrt{2^2}}(|00\rangle + |01\rangle + |10\rangle + |11\rangle)$

In step 3, the series of transformation are took placed known as Grover's iteration. The Amplitude amplification of search key is performed in this iterative step. $O(\sqrt{N})$ number of Grover's iteration step is required. In first Grover's iteration is to call *Quantum Oracle 'O'*. An oracle is basically a black box and the oracle can be written as

$$|x\rangle \xrightarrow{Q_{oracle}} (-1)^{f(x)}|x\rangle, \text{ where } f(x) = 1 \text{ if } |x\rangle \text{ is correct otherwise } f(x) = 0. \qquad (3)$$

Next step in Grover's iteration is diffusion transformation which perform inversion about the average. Transformation of amplitude of each state is performed such that it is far above from average as it was below the average before transformation. The diffusion transformation is carried out with Hadamard transformation denoted by $H^{\otimes n}$ followed by Grover's operator. The Grover's operator can be represented by unitary operator $2|0\rangle\langle 0| - I$. The entire diffusion can represented by $H^{\otimes n}[2|0\rangle\langle 0| - I]H^{\otimes n}$ which can be written as $2|\psi\rangle\langle\psi| - I$ and entire Grover's iteration is denoted by $[2|\psi\rangle\langle\psi| - I]O$ and equals to search key $(x_0)$.

Recently, in 2008, Ahmed Younes proposed constant time Quantum search algorithm using entanglement and phase shifts based on Hamming distance[6]. To make an item using phase shift of $\alpha$ and an oracle $O_{f\alpha}$ is given by

$$O_{f\alpha}|x\rangle = e^{i\alpha f(x)}|x\rangle \quad (4)$$

and to make an item entangled, an oracle $O_{fx}$ is defined as

$$O_{f_x}|x,y\rangle = |x, y \oplus f(x)\rangle \quad (5)$$

In this algorithm both entanglement and phase shift of $\alpha$ is used where an oracle is form of $e^{i\alpha O_f}$ where $O_f$ is defined as

$$O_f|x,0\rangle = |x, f(x)\rangle \quad (6)$$

## 4. Proposed Algorithm

In this proposed quantum search algorithm, the algorithm required 'n' numbers of quantum register of 'n' qubits, where 'n' is required to specify N=2$^n$ numbers of search space and all 'n' qubits are initialized to $|0\rangle$. In this algorithm, a constant number of transformation is took place. At first setup an oracle 'O', black box that will modify the system depending on whether it is in the configuration we are searching for, let $x_s$ is search key. The oracle function is defined system state as shown in equation 7.

$$|\psi_1\rangle = H^{\otimes n}|x_s\rangle \quad (7)$$

and n parallel Hadamard gate is applied over each n- qubit i.e. Hadamard transformation is applied over search space $|x\rangle$, which can be defined by equation 8.

$$|\psi_2\rangle = H^{\otimes n}|x\rangle \quad (8)$$

Subsequently, Apply the Z transformation over $|\psi_2\psi_1\rangle$ and doted by

$$|\psi'_2\rangle|\psi'_1\rangle = (I^{\otimes 2n} \otimes Z)|\psi_2\rangle|\psi_1\rangle \quad (9)$$

After Z transformation output is applied to Hadamard transformation over $|\psi'_2\psi'_1\rangle$ as a result we have

$$|\psi''_2\rangle|\psi''_1\rangle = (H^{\otimes 2n})|\psi'_2\rangle|\psi'_1\rangle \quad (10)$$

Next, we define a quantum interference and function $U_f(x,y) = |x, x \oplus y\rangle$. The quantum state $|\psi''_1\rangle$ and $|\psi''_2\rangle$ applied to function $U_f(x,y)$ where $x, y$ represent $|\psi''_1\rangle$ and $|\psi''_2\rangle$. Finally, the output is measured from lower n-qubit. Result will show the certainty for given search key. The Algorithm is shown bellow

**Algorithm: Quantum Search**

**Input:** Given a list of datasets x known as search space with n-Qubit and Size of search space N= 2$^n$ number and search key $x_s$

**Output:** show the certainty for given search key $x_s$

**Step1:** The oracle defined system state as

$$|\psi_1\rangle = H^{\otimes n}|x_s\rangle = \frac{1}{\sqrt{2^n}} \sum_{i=0}^{2^n-1} (|x_{si}\rangle \otimes |0\rangle)$$

**Step2:** Perform hadamard transform of $|x\rangle$.

$$|\psi_2\rangle = H^{\otimes n}|x\rangle = \frac{1}{\sqrt{2^n}}\sum_{i=0}^{2^n-1}(|x_i\rangle \otimes |0\rangle)$$

**Step3:** Apply Z transformation to perform $180^0$ phase shift over $|\psi_1\rangle$ and $|\psi_2\rangle$.

Then, $|\psi'_1\rangle = (I^{\otimes n}\otimes Z)|\psi_1\rangle$ and $|\psi'_2\rangle = (I^{\otimes n}\otimes Z)|\psi_2\rangle$

**Step 4:** $\forall\ i = 1$ to n

Perform Hadamard transformation over $|\psi'_1\rangle$ and $|\psi'_2\rangle$.

Then, $|\psi''_1\rangle = H^{\otimes n}|\psi'_1\rangle = \frac{1}{\sqrt{2^n}}\sum_{i=0}^{2^n-1}(|\psi'_{1i}\rangle \otimes |0\rangle)$

and $|\psi''_2\rangle = H^{\otimes n}|\psi'_2\rangle = \frac{1}{\sqrt{2^n}}\sum_{i=0}^{2^n-1}(|\psi'_{2i}\rangle \otimes |0\rangle)$

**Step 5:** Apply function $U_f(x,y) = |x, x \oplus y\rangle$ over $|\psi''_1\rangle$ and $|\psi''_2\rangle$

**Step 6:** Measure the lower n-qubit, shows the certainty for given search key $x_s$.

Figure 4, depicted the step wise transformation over search space x and desire search key $x_s$. In this proposed, it is clear that we need not have iteration unlike Grover's iteration. each step of transformation is required constant time. Hence, the running time of proposed quantum search algorithm is O(1) in which can shows the certainty for given search key $x_s$.

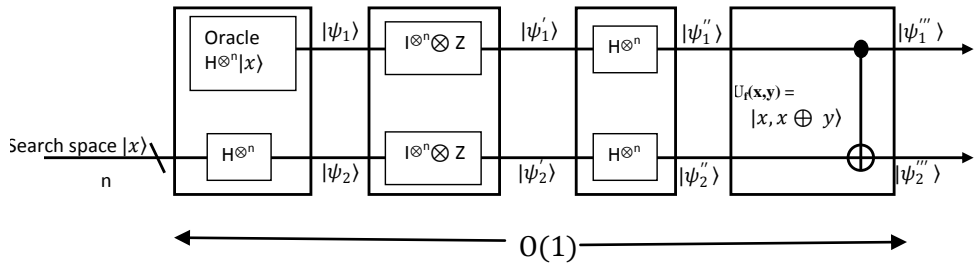

Fig 4 circuit diagram for constant time quantum search algorithm

## 5. Experimental Result

The proposed quantum search algorithm is executed over IBM Quantum Experience (IBMQ) QISKit SDK[1] in python 3.6. let as assume that 2 numbers of quantum register of '2' qubits and $N=2^2$ search space. and search key $|x_s\rangle = |01\rangle$. The corresponding quantum circuit and result is shown in figure 5 and 6 respectively.

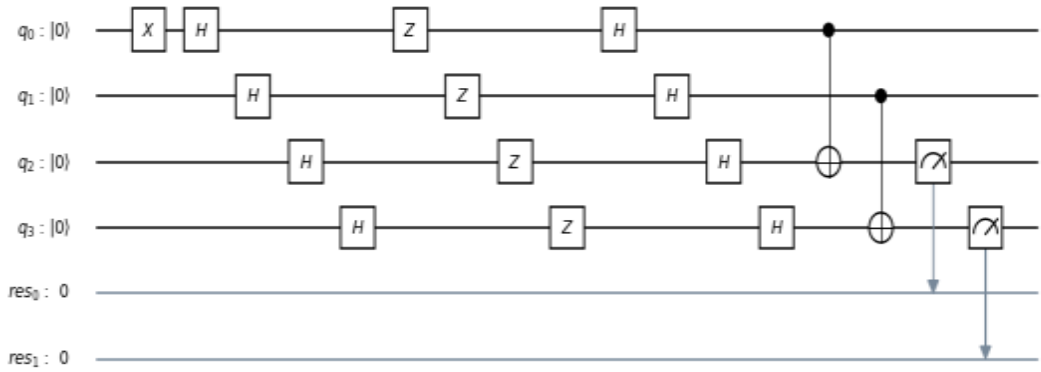

Fig 5 Quantum circuit for constant time search executed in IBMQ QISKit SDK

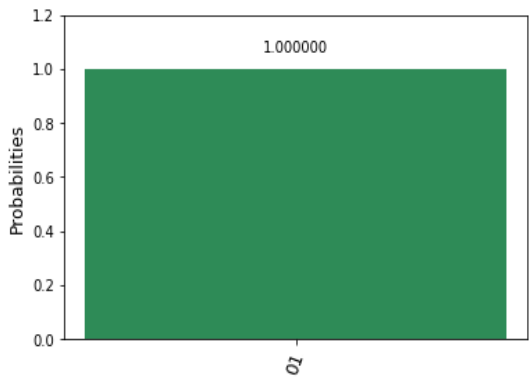

(a)

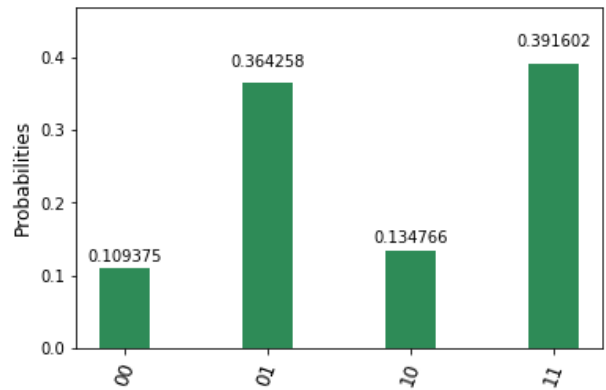

(b)

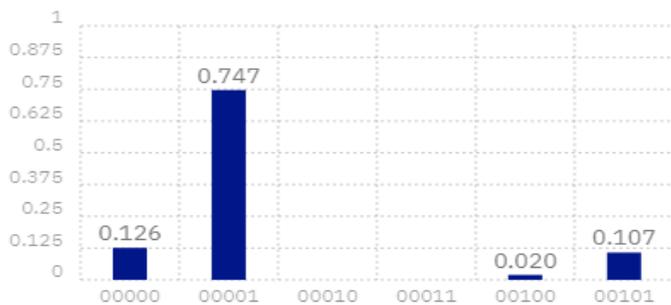

(c)

```
OPENQASM 2.0;              z q[2];
include "qelib1.inc";      z q[3];
qreg q[4];                 h q[0];
creg res[2];               h q[1];
x q[0];                    h q[2];
h q[0];                    h q[3];
h q[1];                    cx q[0],q[2];
h q[2];                    cx q[1],q[3];
h q[3];                    measure q[2] -> res[0];
z q[0];                    measure q[3] -> res[1];
z q[1];
```

(d)

Fig 6. certainty of search key $x_s = 01$ (a) executed in Local backend 'local_qasm_simulator' (b) executed in real chip 'ibmq_16_melbourne' IBMQ, (c) executed in real chip 'ibmqx4' IBMQ and (d) corresponding QASM code for the above quantum circuit.

## 6. Conclusion

Using quantum superposition, set of transformation, quantum interference and $\pi$ phase shift, in quantum computer can search a key $x_s$ from N=$2^n$ search space in constant running time. The proposed quantum algorithm is executed in QISKit SDK local backend 'local_qasm_simulator', real chip 'ibmq_16_melbourne' and 'ibmqx4' IBMQ. Comparative study among local backend and real chip 'ibmq_16_melbourne' and 'ibmqx4' is drawn in terms of probability of certainty of search key $x_s$, shown in figure 6. It shows that execution result using local backend 'local_qasm_simulator' and real chip 'ibmqx4' the probability of finding search key $x_s = 01$ is 1 and 0.747, while executing in real chip 'ibmq_16_melbourne' the probability of finding search key $x_s = 01$ is very less and it is measured 0.364. It is happening due to quantum error and noise. Hence, it can be concluded that quantum search algorithm can run in quantum computer in constant time.

## Acknowledgement

The authors Dr. Kunal Das, Arindam Sadhu are grateful to The SCIENCE & ENGINEERING RESEARCH BOARD (DST-SERB), Govt. of. India, for providing with the grant for accomplishment of the project under the Project FILE NO.CR/2016/000613. and We acknowledge use of the IBM Q experience for this work.